%% file: 0_main.tex
\documentclass[sigconf]{acmart}
\acmSubmissionID{258}

\usepackage[ruled,noend]{algorithm2e} 

\SetAlFnt{\small}
\SetAlCapFnt{\small}
\SetAlCapNameFnt{\small}
\SetAlCapHSkip{0pt}
\usepackage{graphics}
\SetKwInput{KwInput}{Input}                
\SetKwInput{KwOutput}{Output}              
\SetArgSty{textnormal}

\newcommand{\moniker}{ContourCraft}

\copyrightyear{2024}
\acmYear{2024}
\setcopyright{rightsretained}
\acmConference[SIGGRAPH Conference Papers '24]{Special Interest Group on Computer Graphics and Interactive Techniques Conference Conference Papers '24}{July 27-August 1, 2024}{Denver, CO, USA}
\acmBooktitle{Special Interest Group on Computer Graphics and Interactive Techniques Conference Conference Papers '24 (SIGGRAPH Conference Papers '24), July 27-August 1, 2024, Denver, CO, USA}
\acmDOI{10.1145/3641519.3657408}
\acmISBN{979-8-4007-0525-0/24/07}

\begin{document}
\title{\moniker{}: Learning to Resolve Intersections in Neural Multi-Garment Simulations}

\author{Artur Grigorev}
\orcid{0000-0002-6999-2162}
\affiliation{%
 \institution{ETH Zurich}
 \country{Switzerland}}
\affiliation{%
 \institution{Max Planck Institue for Intelligent Systems}
 \country{Germany}}
\email{agrigorev@ethz.ch}

\author{Giorgio Becherini}
\orcid{0009-0005-8770-8144}
\affiliation{%
 \institution{Max Planck Institue for Intelligent Systems}
 \country{Germany}}
\email{giorgio.becherini@tuebingen.mpg.de}

\author{Michael J. Black}
\orcid{0000-0002-5068-3474}
\affiliation{%
 \institution{Max Planck Institue for Intelligent Systems}
 \country{Germany}}
\email{black@tuebingen.mpg.de}

\author{Otmar Hilliges}
\orcid{0000-0002-5068-3474}
\affiliation{%
 \institution{ETH Zurich}
 \country{Switzerland}}
\email{otmar.hilliges@inf.ethz.ch}

\author{Bernhard Thomaszewski}
\orcid{0000-0002-5068-3474}
\affiliation{%
 \institution{ETH Zurich}
 \country{Switzerland}}
\email{bthomasz@ethz.ch}

\definecolor{orange}{HTML}{FF5900}
\newcommand{\todo}[1]{\textcolor{orange}{#1}}
\input{include/figure_teaser}

\begin{abstract}
Learning-based approaches to cloth simulation have started to show their potential in recent years. However, handling collisions and intersections in neural simulations remains a largely unsolved problem.
In this work, we present \moniker{}, a learning-based solution for handling intersections in neural cloth simulations. 
Unlike conventional approaches that critically rely on intersection-free inputs, \moniker{} robustly recovers from intersections introduced through missed collisions, self-penetrating bodies, or errors in manually designed multi-layer outfits. 
The technical core of \moniker{} is a novel intersection contour loss that penalizes interpenetrations and encourages rapid resolution thereof. 
We integrate our intersection loss with a collision-avoiding repulsion objective into a neural cloth simulation method based on graph neural networks (GNNs).
We demonstrate our method's ability across a challenging set of diverse multi-layer outfits under dynamic human motions. 
Our extensive analysis indicates that \moniker{} significantly improves collision handling for learned simulation and produces visually compelling results.

\end{abstract}
\maketitle

\input{1_Introduction}
\input{2_Related_Work}
\input{3_Background}
\input{4_Method}
\input{5_Results}

\input{6_Conclusion}

\input{7_Limitations}

\bibliographystyle{ACM-Reference-Format}
\bibliography{main}

\input{include/figure_oneseqplot}

\appendix
\input{91_SM_Derivation}
\input{92_outfits}
\input{93_implementation}

\end{document}

%% file: include/figure_teaser.tex
\begin{teaserfigure}
    \centering
    \includegraphics[width=\linewidth]{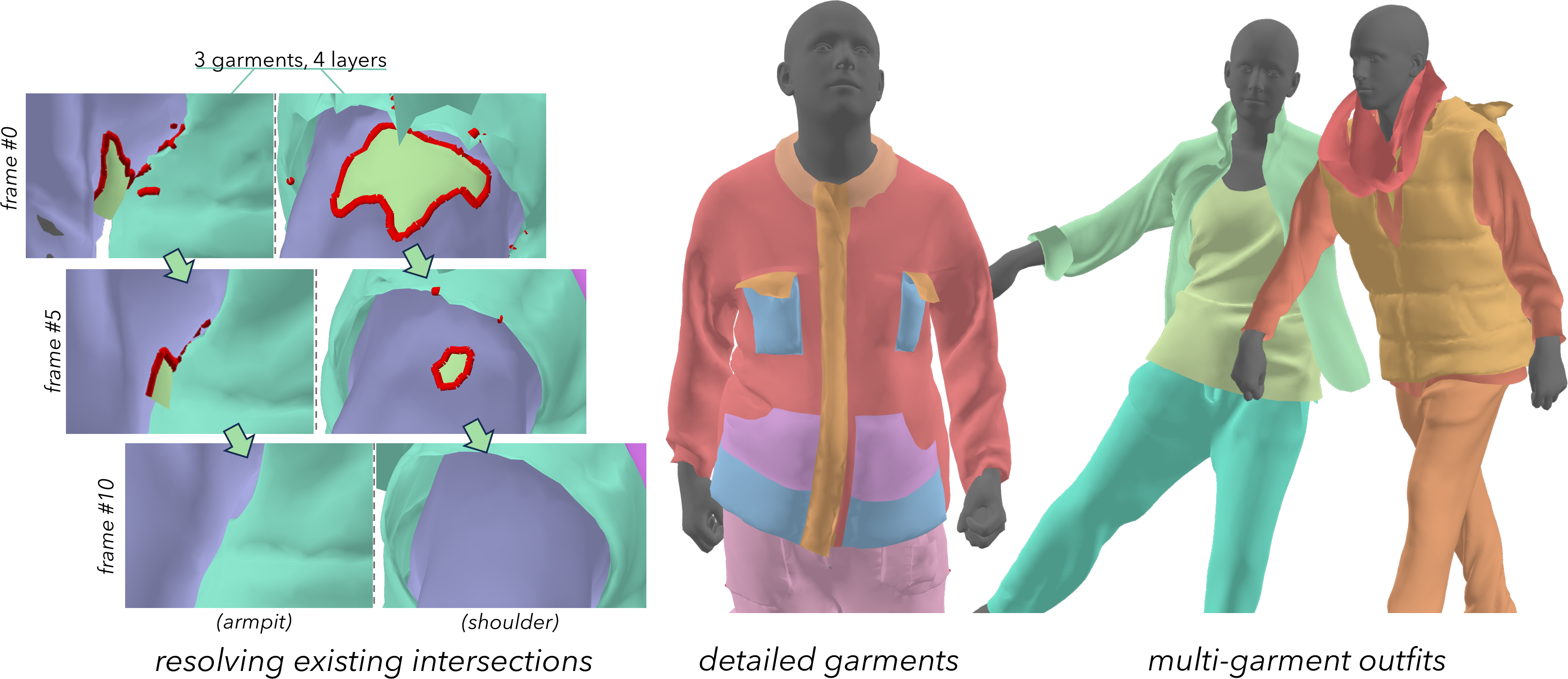}
      \caption{
      We present a novel graph neural network-based approach to learned simulation of multilayered garments. Its key component is an Intersection Contour objective term that encourages resolution of existing cloth--cloth intersections. 
      Even when initialized with intersecting meshes, our approach resolves
      penetrations \textit{(left)}, thus opening the door to learning-based simulation of detailed multi-layer garments \textit{(middle)} and multi-garment outfits \textit{(right)}.}
      \label{fig:teaser}
\end{teaserfigure}

%% file: 1_Introduction.tex
\section{Introduction}

Garment simulation plays a crucial role in video games, animated movies, special effects, fashion design, and many other applications involving digital humans. 
While conventional methods produce compelling results for complex garments, computational demands for high-quality animations can be significant.
Neural simulation methods have emerged as a promising alternative, but a common limitation is their inability to reliably prevent or handle garment intersections. As a result, complex garments and multilayer outfits can exhibit a substantial number of missed intersections, leading to visually disturbing artifacts and overall implausible motion.

Conventional methods, in contrast, aim to maintain intersection-free garments throughout the simulation. 
With this condition met in the initial state, these methods iteratively resolve all penetrations occurring from one time step to the next. 
This strategy's critical disadvantage is that it relies on an intersection-free state. 
Collision response will try to maintain any missed or pre-existing collision, which can compromise entire animation sequences. Unfortunately, intersections due to garment pre-positioning, user interaction, or self-intersecting body motion cannot always be avoided. 
 
In this work, we propose a novel approach for handling intersections in neural cloth simulations (see Fig.~\ref{fig:teaser}). Instead of \textit{avoiding} intersections at all costs, we propose a mechanism for \textit{recovering} from existing intersections. We draw inspiration from previous work by Volino and Magnenat-Thalmann \shortcite{ICM}, who propose Intersection Contour Minimization (ICM) as a means of resolving penetrations. 
As our key contribution,  we introduce a variational formulation of ICM that allows for direct integration into learning-based methods. 
The contour loss works in combination with a repulsion term to \textit{avoid and resolve} cloth intersections.
We furthermore use intersection contours to identify penetrating regions, which allows us to apply targeted repulsion forces without preventing intersections from resolving.
We integrate our new strategy with an existing neural simulation method based on GNNs and unsupervised training \cite{hood}.

We demonstrate the effectiveness of our method by simulating a diverse set of complex multi-layer outfits for dynamic human motion. In particular, we show that our method rapidly recovers from heavily intersecting configurations of complex multi-layer outfits. Intersections introduced during fast or self-penetrating motion of the underlying body are also reliably resolved. 
All results can be reproduced using our publicly available source code\footnote{URL will be disclosed upon acceptance.}.

%% file: 2_Related_Work.tex
\section{Related Work}
Most related to our work is research on collision handling for garment simulation and learning-based approaches to cloth simulation.

\paragraph{Collision Handling for Garment Simulation}
Detecting and handling collisions has been a central focus of computer animation for several decades \cite{teschner2005collision,baraff2003untangling}. 
Bridson et al.~\shortcite{bridson2002robust} presented a three-tier collision-handling pipeline that uses impulses, i.e., velocity corrections, to prevent imminent collisions. 
The first stage applies repelling impulses that push primitives apart that are too close together. 
The second stage uses continuous collision detection (CCD) and corresponding impulses to resolve intersections occurring during time steps. 
Any collisions that remain after the second stage are treated using rigid impact zones, which cancel relative velocities between regions spanning multiple collision primitives. This impulse-based collision handling framework has been widely adopted and extended in various ways~\cite{harmon2008robust}.

Whereas the approach by Bridson et al.~\shortcite{bridson2002robust} decouples time integration and collision handling, recent work explores more closely coupled treatments. 
Harmon et al.~\shortcite{Harmon2009} describe an approach for asynchronous time stepping of contact dynamics that guarantees robustness at the expense of significantly increased computation times. 
More recently, IPC~\cite{li2020incremental} provides robust and fully implicit treatment of contact dynamics using smoothly clamped log barrier functions. 
While IPC can generate compelling animations with complex and challenging self-collisions, computation times are on the order of minutes per frame. 

Another line of work~\cite{tang2011collision,tang2013gpu,tang2018cloth,tang2018pscc,wu2020safe,wang2017efficient,wang2021gpu} aims to accelerate the collision handling process by developing efficient and highly parallelizable algorithms that leverage the processing power of GPUs.
These works mainly focus on optimizing the CCD step as it is the most time-consuming part of the collision handling process.
The modifications to CCD include efficient hashing to accelerate the traversal of bounding volume hierarchies (BVH)~\cite{karras2012maximizing,tang2018pscc}, tests that enable parts of the BVH to be skipped \cite{wang2017efficient}, and incremental collision detection that handles only triangles affected by collision-response steps \cite{wang2017efficient}.
While being much faster than non-optimized CPU algorithms, these approaches still perform the same sequence of steps used in the classical method of~\cite{bridson2002robust}.
Wu et al.~\shortcite{wu2020safe} and Wang \shortcite{wang2021gpu} propose to substitute continuous collision constraints with a set of discrete ones that greatly improve the simulation speed.
However, to do that they rely on a regular grid-like mesh structure maintained by dynamic re-meshing.

All of the above method require intersection-free geometry as input and follow the conventional strategy of detecting and resolving new collisions between time frames. 
Because of that, they struggle to recover from intersections that either exist in the initial geometry or occur during the simulation.
The same mechanisms that prevent intersections from occurring also prevent them from being resolved.

Few works have tackled the problem of resolving existing intersections.
Baraff et al.~\shortcite{baraff2003untangling} detect connected components separated from the rest of the mesh by a closed contour of penetrations.
To resolve these intersections, their method applies attractive forces to the pairs of such components.
While intersections with closed contours can be resolved in this way, their method does not address open-contour intersections such as penetrating pairs of mesh boundaries.
Intersection Contour Minimization (ICM) \cite{ICM} aims to resolve cloth intersections in the general case by minimizing the length of intersection contours.
For each pair of intersecting triangles, ICM computes the gradient of intersection length with respect to the mesh vertices involved in the corresponding collisions.
It then iteratively applies local displacements in the negative direction of the gradients such as to resolve the intersections.
While several subsequent works have adopted and extended this idea~\cite{zhong2009fast,ye2012intersection,ye2015fast,ye2017unified,cha2020tanglement}, we are the first to explore Intersection Contour Minimization in the context of learned garment simulation.
Rather than applying intersection-resolving forces, we train a GNN with an objective term that penalizes the total length of the intersections. In this way, our model learns to resolve intersections in an unsupervised manner.

\paragraph{Learning-Based Cloth Simulation}
To avoid the high computational cost of physics-based cloth simulation, a recent stream of work has started to explore learning-based techniques for this task.
Learned deformation models are commonly used to model garment behavior~\cite{guan2012drape,santesteban2019learning,santesteban2021self, santesteban2022snug, bertiche2022neural}. 
These models predict garment deformations based on the pose and shape of the underlying body model.
While offering fast inference due to their relatively small network sizes, they have to be retrained separately for each garment and cannot generalize to unseen ones.
As an alternative approach, MeshGraphNets~\cite{meshgraphnets} uses graph neural networks to learn mesh-based simulations from examples.
Trained on cloth simulation data, MeshGraphNets learns to handle cloth--cloth interactions in the process.
Taking this approach one step further, HOOD~\cite{hood} learns the dynamic behavior of garments and their interaction with the human body.
The unsupervised physics-guided training, inspired by PBNS~\cite{bertiche2020pbns} and SNUG~\cite{santesteban2022snug}, alleviates the need for curated simulation data.
However, HOOD is unable to model cloth--cloth interactions and self-collisions, leading to unrealistic results for multi-layer outfits.
LayersNet~\cite{layersnet} employs a transformer-based approach to model multi-layer outfits.
However, it requires a large dataset of physically simulated ground-truth data to train.

Several recent learning-based approaches proposed strategies to address garment collisions. 
For instance, Repulsive Force Unit (ReFU)~\cite{tan2022repulsive} operates as a plug-in layer that processes the human body as a signed distance field (SDF), pushing garment nodes and edges outside of it.
Implicit Untangling~\cite{buffet2019implicit} and ULNeF~\cite{santesteban2022ulnef} both use implicit representations for garment untangling. 
Given a set of interpenetrating garments in a canonical pose, these techniques resolve penetrations by arranging garments in a specific order, i.e., from innermost to outermost.
However, these methods only consider static garments in a single canonical pose and do not handle dynamic collisions.
PBNS~\cite{bertiche2020pbns} prevents inter-garment penetrations in various poses by applying a collision penalty between layer-ordered garments during training.
For each garment, PBNS finds and penalizes intersections with the body and lower-level garments.
However, PBNS requires training a separate model for each set of garments and does not account for intra-garment collisions, making it unsuitable for modeling multi-layer outfits. 
In summary, while some learning-based methods account for penetrations between different garments, none of them addresses the problem of preventing and resolving garment self-penetrations in dynamic scenes. 
Our approach addresses this significantly more challenging problem by training a GNN to prevent and resolve garment intersections.

Another recent stream of work investigates using neural networks to learn nonlinear subspaces for rapid simulation \cite{Holden19Cloth,Fulton:LSD:2018, Shen21HighOrder,Sharp23Datafree,Wang24Neural}. A notable example in this context is the method by Romero et al.~\shortcite{Romero21Learning} that learns contact corrections for handle-based subspace dynamics. While these methods can yield substantial accelerations compared to full space simulation, the networks are trained for a given input mesh and do not generalize to new input geometry. 

%% file: 3_Background.tex
\vspace{-5pt}
\section{Background: Learned Garment Simulation}

MeshGraphNets~\cite{meshgraphnets}  approaches the task of cloth simulation with the help of graph-based neural networks.
HOOD~\cite{hood} takes this approach one step further and demonstrates how such a model can be trained in an unsupervised physics-guided fashion to predict the dynamics of garments and their interactions with the human body.
HOOD takes as input a graph containing nodes $V^G$ and edges $E^G$ of the garment mesh augmented with so-called \textit{body edges}, $E^B$, that connect the garment vertices to the nodes of the body mesh $V^B$.
Each node and edge of the input mesh is endowed with a corresponding feature vector that describes properties such as nodal mass, velocity, and rest length of the mesh edge.
The feature vectors are then mapped into latent space, updated through several message-passing steps, and finally decoded into nodal accelerations.
The complete process can be written as
\begin{gather}
    \hat{A} = f_\theta (V^G, V^B, E^G, E^B)\ ,
\end{gather}
where $\hat{A}$ are the predicted accelerations for each garment node and $f_\theta$ is the GNN.

The network is trained in an unsupervised manner with the objective function consisting of a set of energies produced by stretching, bending, inertia, and other physical phenomena,
\begin{gather}
    \mathcal{L}_{\mathrm{HOOD}} = \mathcal{L}_{\mathrm{bending}} + \mathcal{L}_{\mathrm{stretching}} + \mathcal{L}_{\mathrm{gravity}} + \\
    + \mathcal{L}_{\mathrm{friction}} + \mathcal{L}_{\mathrm{collision}}^{\mathrm{body}} + \mathcal{L}_{\mathrm{inertia}} \ .
\end{gather}
Building on an optimization-based variant of implicit Euler integration \cite{Martin11EBEM}, the model is thus encouraged to predict accelerations that balance kinetic and potential energy in a robust and physically plausible way.
This method's main drawback is that it completely ignores cloth--cloth interactions, rendering it unsuitable for modeling detailed multi-layer garments and multi-garment outfits. 
We address this limitation by extending HOOD with an intersection contour loss term and supplying it with information about cloth interactions.

%% file: 4_Method.tex
\section{Method}
\label{sec:method}
Our method builds on the concept of intersection contours \cite{ICM}.
We first use intersection contours to differentiate between repulsive and non-repulsive cloth--cloth interactions.
Then, we define intersection contour length as an additional loss term to encourage the model to resolve existing intersections.

To evaluate the advantages of our method, we conduct a detailed ablation study in which we progressively modify the GNN-based approach by Grigorev et al.~\shortcite{hood}. 
In this section, we describe and explain the motivation behind each modification.

\subsection{Conventional Collision Handling with GNNs}
To establish a meaningful baseline, we first develop a direct implementation of conventional collision handling for GNNs. 
To this end, we augment the GNN model with correspondences between spatially close pieces of cloth and include a new repulsion term into its loss function.
This enables the GNN to prevent the majority of penetrations.

\subsubsection{Cloth--Cloth Correspondences.} 
Following MeshGraphNets \cite{meshgraphnets}, we expand the input graph to incorporate cloth--cloth correspondences. 
To identify these correspondences, we detect all face--node pairs for which (\textit{a}) the projection of the node along the face normals falls inside the face and (\textit{b}) the corresponding distance is smaller than a given threshold $\epsilon$ (see Algorithm \ref{algo:ccc}).
Condition (\textit{a}) limits the number of correspondences in the graph and thus accelerates inference. 
For each such pair,
we add a \textit{"world edge"} to the graph connecting the node to the closest triangle vertex from the pair.

The forward pass of the model is formulated as
\begin{gather}
    \hat{A} = f_\theta (V^G, V^B, E^G, E^B, E^W)\ ,
\end{gather}
where $E^W$ is a set of \textit{world} edges.

\subsubsection{Repulsion Loss.} 
To make use of the cloth--cloth correspondences, we augment the training objective with a new repulsion term,
\begin{gather}
\mathcal{L}_{repulsion} = \sum_i max(\xi - d^{(i)}_{curr} \cdot sign(d^{(i)}_{prev}), 0)^3 \ ,
\end{gather}
 where $d_{curr}^{(i)}$ and $d_{prev}^{(i)}$ are distances between the node $v^{(i)}$ and the corresponding face $f^{(i)}$ in the current and previous frames respectively,
\begin{gather}
d^{(i)} = ((v^{(i)} - f^{(i)}) \cdot \vec{n}^{(i)})\ .
\end{gather}
The repulsion term penalizes face--node pairs in which the node either crosses the face or comes closer to it than a given repulsion threshold $\xi$, corresponding to the fabric's thickness (we use $\xi=1mm$).

Since in this version, all cloth interactions are penalized by the repulsive loss, we refer to it as \textit{"only repulsive"}.
Its full objective term is
\begin{gather}
\mathcal{L}^{\mathrm{only}}_{\mathrm{repulsive}} = \mathcal{L}_{\mathrm{HOOD}} + \mathcal{L}_{\mathrm{repulsion}}.
\end{gather}

These two modifications enable the model to prevent most penetrations, but they do come with two major drawbacks.
First, the updated model can still miss collisions.
Therefore, a fail-safe method based on conventional, non-learned collision handling is still required for realistic simulations. 
Second, such a method would still struggle to recover from existing penetrations---the mechanisms designed to prevent self-intersections from occurring also prevent their resolution. 

\subsection{Repulsive and non-repulsive interactions}
To resolve existing self-intersections we must distinguish between two types of cloth--cloth interactions.
In an intersection-free state, spatially close regions of cloth need to be repelled from each other.
However, for existing intersections, such a response might prevent them from being resolved.

We distinguish between repulsive and non-repulsive interactions in the following way.
First, we run discrete collision detection (DCD) to determine all penetrating triangle pairs. 
We then combine sequences of adjacent penetrations into intersection contours.
Each contour may be either open or closed.
The latter splits the surface into two disconnected parts.
In this case, we consider the smaller part to lie inside the contour, and the larger one to lie outside.
To handle cases with nested closed contours, we only keep the outermost ones.
Finally, we mark as \textit{"non-repulsive"} the interactions where at least one node is either (a) part of an open contour or (b) lies inside a closed contour.
See Fig. \ref{fig:node_types} for illustration.

The distinction between repulsive and non-repulsive interactions translates into the corresponding world edges in the input graph.
This leads to an updated expression for nodal accelerations,
\begin{gather}
    \hat{A} = f_\theta (V^G, V^B, E^G, E^B, E^W_R, E^W_{NR})\ ,
\end{gather}
where $E^W_R$ and $E^W_{NR}$ are world edges corresponding to repulsive and non-repulsive interactions, respectively.

\input{include/figure_nodetypes}
\subsection{Na\"ive baseline}
The simplest way of dealing with intersections is to ignore them. 
That is, to ignore those cloth--cloth interactions that participate in any penetration.
We implement this strategy as an ablated model (\textit{"w/o IC loss"}) where all such correspondences are omitted from the graph.
In this way, the model does not prevent existing intersections from being resolved but does not try to purposefully recover from them.

\subsection{Intersection Contour Loss}
To train the model to resolve cloth intersections we employ a simple but effective objective term that we refer to as \textit{Intersection Contour Loss} ($\mathcal{L}_{IC}$).
This term is based on the lengths of all triangle--triangle intersections.
Here we describe the process of computing this term in detail.

Each triangle--triangle intersection detected by DCD contains two edge--triangle intersections.
For each of them, we can first find the relative coordinate $s$ of the intersection point on the edge, 
\begin{gather}
s = \frac{(x_{\Delta}\cdot \vec{n}) - (x_0\cdot\vec{n}) }{(x_1-x_0)\cdot\vec{n}} \ ,
\end{gather}
where $x_0$ and $x_1$ are vertices of the intersecting edge, $x_{\Delta}$ is a point inside the triangle, and $\vec{n}$ is the triangle's normal vector. 
The coordinates of the intersection point are computed as
\begin{gather}
p_{I^j_i} = dg(x_0) + s \cdot dg(x_1-x_0) \ ,
\end{gather}
where $i$ is the index of the intersecting triangle pair and $j$ is the index of the edge--triangle intersection within it.
The function $dg()$ indicates that we do not use the partial derivative of this term w.r.t. its arguments when computing the gradient during training (see Section \ref{sec:icgrad} for details). 
Finally, the loss term is computed as a sum of the squared lengths of all intersections,
\begin{gather}
\mathcal{L}_{\mathrm{IC}} = \sum_i \| p_{I_i^0} - p_{I_i^1}\|^2 \ .
\end{gather}
The full objective function is thus
\begin{gather}
\mathcal{L}_{\mathrm{ours}} = \mathcal{L}_{\mathrm{HOOD}} + \lambda_1\mathcal{L}_{\mathrm{repulsion}} + \lambda_2\mathcal{L}_{\mathrm{IC}},
\end{gather}
where $\lambda_1$ and $\lambda_2$ are weighting coefficients.
In this setting, the repulsive correspondences are explicitly penalized by the repulsion loss, while non-repulsive ones are penalized implicitly by the Intersection Contour Loss.

\subsection{Intersection Contour Gradient}
\label{sec:icgrad}
While the intersection contour objective is straightforward to integrate with the physics-based loss, directly using it during training leads to sub-optimal results.
Fig. \ref{fig:icgrad} shows an example in which the gradient of length for a single contour segment is decomposed into its partial derivatives. 
One of the components reduces length by distorting the triangle, whereas the other one achieves length reduction through translational motion. Please refer to the Supplemental Material for details.
While nominally reducing length, we empirically observe that the distortional component is generally not helpful for resolving intersections as it induces compression in the fabric and provokes strong reaction forces. 
For this reason, we use only the partial gradient corresponding to quasi-rigid translation during training. 
This strategy successfully reduces distortion artifacts and, as our statistical analysis shows (Table \ref{tab:ablation}, Fig. \ref{fig:penat50}), leads to fewer intersections overall.
In our experiments we refer to the model where both components are used as \textit{"full gradient."}

\input{include/figure_icgrad}

\input{include/algo_cc}
\input{include/algo_buildGraph}

\subsection{Building the input graph}
Algorithm \ref{algo:buildGraph} shows the overall process of building the input graph for the GNN.
Similar to HOOD, we initialize from a graph of the mesh edges (\textit{buildGraph}) and then augment them with edges to the closest body nodes (\textit{addBodyEdges}).
We then detect intersecting triangle pairs within the outfits with DCD and combine adjacent intersections into intersection contours (\textit{makeContours}), removing contours that are enclosed by others (\textit{removeNested}).
Finally, we find face--node correspondences in world coordinates (\textit{findClothCorrespondences}, Algorithm \ref{algo:ccc}), classify them into repulsive and non-repulsive, and add corresponding world edges to the graph.

\subsection{Training process}
Our method is trained to autoregressively predict nodal accelerations that generate realistic garment motions and also prevent and resolve cloth self-intersections.
We split the training process into three stages.

The first stage follows exactly the training process of HOOD~\cite{hood}. 
Its goal is to train a model to realistically mimic the physical behavior while ignoring cloth--cloth interactions.
To this end, we exclude the corresponding interactions from the input graph and train it with the original objective function, $\mathcal{L}_{\mathrm{HOOD}}$.

The second stage aims to teach the model to \textit{prevent} cloth self-intersections. 
For this purpose, we remove all self-intersections from the garment mesh in the initial frame and then resolve all collisions happening between frames using the method by Bridson et al. \shortcite{bridson2002robust}. 
If the algorithm fails to resolve all collisions in a given number of iterations, we stop the simulation of the current sequence and move on to the next one in the dataset.
Note that the algorithm by Bridson et al. ~\shortcite{bridson2002robust} is only used to maintain intersection-free geometry in this stage of the training process and is not part of the loss function.
In this stage, all cloth--cloth interactions are treated as \textit{repulsive}, so the objective function is $\mathcal{L}_{\textrm{HOOD}} + \mathcal{L}_{\textrm{repulsion}}$.

The third stage is designed to teach the model to \textit{resolve} cloth intersections.
Here, we again initialize from potentially self-intersecting geometry without resorting to \cite{bridson2002robust} and classify the cloth--cloth interactions into \textit{repulsive} and \textit{non-repulsive} ones (see Fig. \ref{fig:node_types}).
During this stage, we use the full objective function, $\mathcal{L}_{\textrm{HOOD}} + \mathcal{L}_{\textrm{repulsion}} + \mathcal{L}_{\textrm{IC}}$.
For more details on implementation and the training process, please refer to the supplemental material.

%% file: include/figure_nodetypes.tex
\begin{figure}[ht]
  \centerline{  \includegraphics[width=\linewidth]{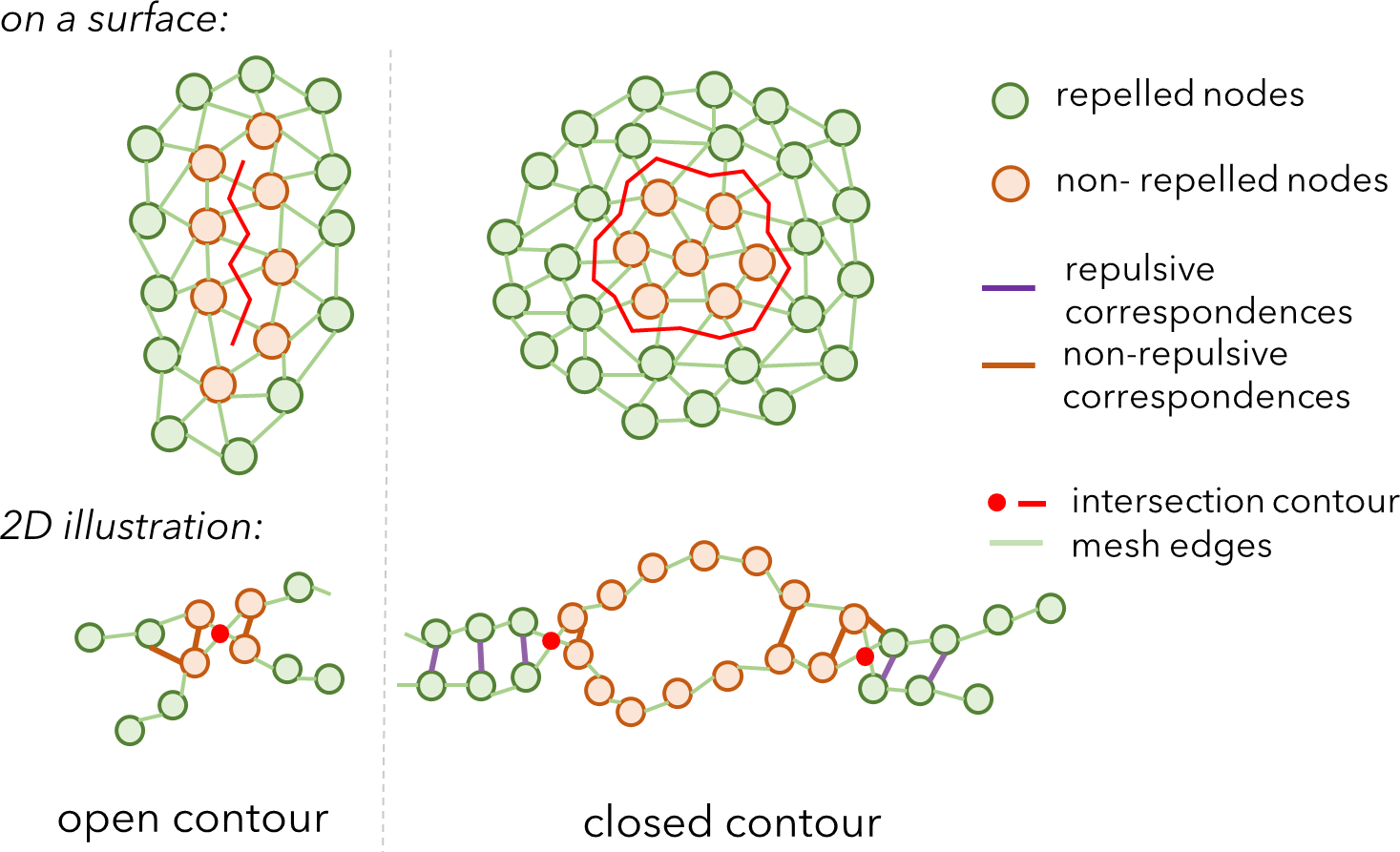}}
  \caption{We distinguish between two types of garment nodes. Repelled nodes are those that either do not participate in the penetrations or lie outside a closed contour. Non-repelled nodes are those that are either part of an open contour or lie inside a closed one. If a triangle-node correspondence only contains repulsive nodes, we define this correspondence as repulsive and apply a repulsion loss to it, otherwise it is non-repulsive.
  \vspace{-10pt}
  }
  \label{fig:node_types}
\end{figure}

%% file: include/figure_icgrad.tex
\begin{figure}[ht]
  \centerline{  \includegraphics[width=\linewidth]{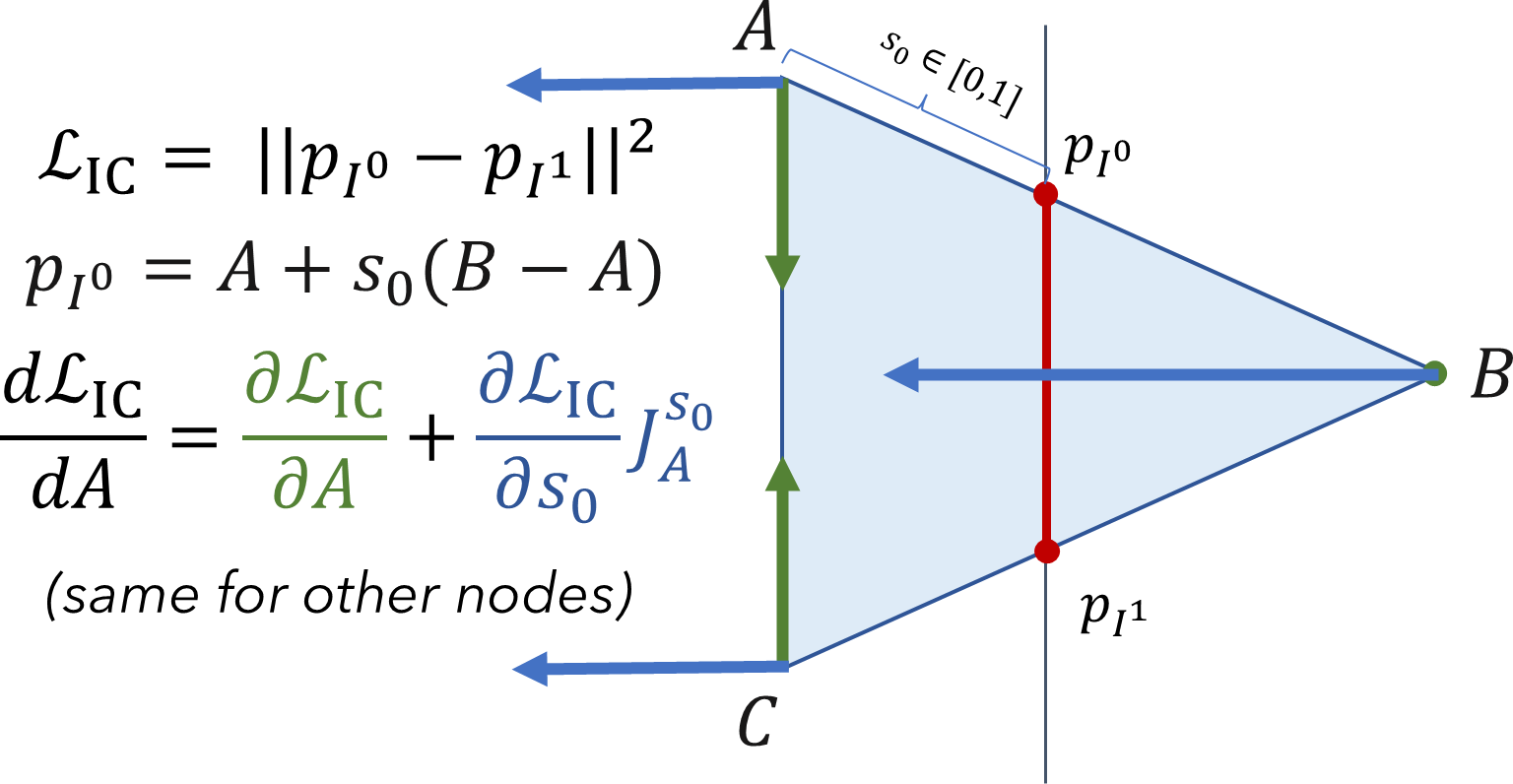}}
  \caption{ A face (blue triangle) intersects a perpendicular plane (vertical line with intersecting segment shown in red). Green arrows show the negative partial gradients of the contour loss $\mathcal{L}_{IC}$ w.r.t.~the triangle nodes. Blue arrows indicate negative partial gradients w.r.t.~the coordinate of the intersection point $s_j$. The former gradient (green) squeezes the triangle to decrease the contour length, while the latter (blue) moves it along the plane's normal direction to resolve the intersection. We only use the gradient w.r.t. $s_j$ in our training. $J_A^{s_0}$ is the Jacobian of $s_0$ w.r.t.~$A$.
  \vspace{-5pt}
  }
  \label{fig:icgrad}
\end{figure}


%% file: include/algo_cc.tex
\begin{algorithm}
\SetKwData{correspondences}{correspondences}

\SetKwFunction{distance}{distance}
\SetKwFunction{append}{append}

\DontPrintSemicolon
\caption{$findClothCorrespondences$}
\label{algo:ccc}
    \KwInput{Outfit mesh with vertices $V$ and faces $F$}
    \correspondences $\leftarrow \{\emptyset\}$

    \ForEach{$v \in V$}{
        \ForEach{$f \in F$} {
            \uIf{$v \in f$} {\textbf{continue}}
        
            \uIf{\distance{$v, f$} $> \epsilon$} {\textbf{continue}}

            $v_{proj} \leftarrow$ \emph{projection of} $v$ \emph{onto} $f$

            \uIf{$v_{proj}$ \emph{inside} $f$} {
                \correspondences.\append{($v, f$)}
            }
        }
    }
    \Return \correspondences
\end{algorithm}

%% file: include/algo_buildGraph.tex
\begin{algorithm}

\SetKwData{Graph}{G}
\SetKwData{Penetrations}{penetrations}
\SetKwData{Contours}{contours}
\SetKwData{ClothCorrespondences}{clothCorrespondences}
\SetKwData{C}{C}
\SetKwData{True}{True}
\SetKwData{False}{False}
\SetKwData{isRepulsive}{isRepulsive}
\SetKwData{contour}{contour}

\SetKwFunction{BuildGraph}{buildGraph}
\SetKwFunction{addBodyEdges}{addBodyEdges}
\SetKwFunction{DCD}{DCD}
\SetKwFunction{makeContours}{makeContours}
\SetKwFunction{removeNested}{removeNested}
\SetKwFunction{buildInputGraph}{buildInputGraph}
\SetKwFunction{findClothCorrespondences}{findClothCorrespondences}
\SetKwFunction{isClosed}{isClosed}
\SetKwFunction{addWorldEdge}{addWorldEdge}

\caption{$buildInputGraph$}
\label{algo:buildGraph}
    \KwInput{Outfit mesh with vertices $V_g$ and faces $F_g$}
    \KwInput{Body mesh with vertices $V_b$ and faces $F_b$}

    \Graph$\leftarrow$\BuildGraph{$V_g, F_g$}
    \tcp*[f]{\emph{build graph from mesh edges}}
    
    \Graph$\leftarrow$\addBodyEdges{\Graph, $V_b$}
    \tcp*[f]{\emph{add "body edges" as in HOOD}}

    \Penetrations $\leftarrow$ \DCD{$V_g, F_g$}

    \Contours$\leftarrow$\makeContours{\Penetrations}

    \Contours$\leftarrow$\removeNested{\Contours}

    \ClothCorrespondences$\leftarrow$\findClothCorrespondences{$V_g, F_g$}

    \ForEach{\C $\in$ \ClothCorrespondences} {
        \isRepulsive$\leftarrow$\True
        
        \ForEach(comment){$v \in C$} {
            \ForEach{\contour $\in$ \Contours} {
                \If{$v$ \emph{is part of or enclosed within} \contour}{
                    \If{\isClosed{\contour}}{
                        \If{$v$ inside \contour} {
                        \isRepulsive$\leftarrow$\False
                            
                        }
                    } 
                    \Else{
                        \isRepulsive$\leftarrow$\False
                    }
                }
            }
            
        }

        \Graph$\leftarrow$\addWorldEdge{\Graph, \C, \isRepulsive}
    }
    \Return \Graph
\end{algorithm}

%% file: 5_Results.tex
\section{Results}
\label{sec:results}

\input{include/figure_resize}

\subsection{Pose sequences and garment meshes}
Our method is trained in an unsupervised physically guided fashion without the need for ground-truth physical simulations.
The only data necessary for training and inference are human body pose sequences and static outfit meshes.

For training we use the same pose sequences from AMASS~\cite{mahmood2019amass} that were used in \citeN{santesteban2021self,santesteban2022snug,hood}. 
We train the model using a combination of simple one-layer garments used in \citeN{santesteban2022snug,hood} and three multi-layer outfits from the BEDLAM dataset~\cite{bedlam}.

For our experiments, we use the same eight pose sequences from AMASS as Grigorev et al. \shortcite{hood} and model five multi-layer outfits from BEDLAM: two outfits with two garments, two outfits with three garments, and one with five garments. Please see the supplementary material for more details.
These outfits were not seen by the model during training. 
In total this amounts to 40 validation sequences.
We start the simulations from the first frame of the pose sequence without initialization steps that interpolate between the T-pose and the first frame pose.
To initialize the garment geometry, we use simple linear blend skinning with the skinning weights collected from the body model.
This means the initial garment geometries may have severe self-penetrations, which our method can resolve (see Fig. \ref{fig:teaser}, left).

To illustrate the practicality of our method, we show rendered sequences of our clothing simulations using the Unreal engine. 
Please refer to the supplementary video for more results.

\subsection{Ablation study}
We compare our method to three ablated models introduced in Section \ref{sec:method}.
To reiterate, in  \textit{"only repulsive"} we consider all cloth-cloth interactions as repulsive; 
in \textit{"w/o IC loss"} we distinguish between repulsive and non-repulsive interactions, but do not apply any supervision to the non-repulsive ones;
in \textit{"full gradient"} we use the Intersection Contour loss with full gradient (see Fig. \ref{fig:icgrad}).

We simulate all 40 validation sequences with each of the ablated models and compare their performance in terms of the average number of intersecting triangle pairs in each frame.
Table \ref{tab:ablation} shows that each modification we introduce significantly reduces the number of intersections in simulated sequences.

\input{include/figure_penat50}

Fig. \ref{fig:penat50} shows how well each ablated model resolves the intersections in the initial geometry.
For each frame, it plots the fraction of the remaining intersections relative to the initial geometry.
This information is aggregated across all 40 validation sequences.
Fig. \ref{fig:oneseqplot} plots the number of intersections for each frame of a single sequence.
Both of these plots demonstrate the improvements in collision handling by each modification.

\input{include/table_statistics}

\subsection{Perceptual study}
Additionally, we perform a perceptual study to assess the realism of \moniker{} results as perceived by human subjects. We compare \moniker{} to CLO3D~\cite{clo3d}, a commercial software for garment design and modeling, and two baseline methods: HOOD~\shortcite{hood} and linear blend skinning (LBS)

The study consists of two parts. In the first part, the participants were asked to rate the realism of the simulations on a scale from 1 to 5. CLO3D scored on average \textbf{4.29}, \moniker{} scored \textbf{4.15}, LBS achieved \textbf{3.04}, and HOOD was rated \textbf{2.78}.
In the second part, the two baselines and \moniker{} were compared to CLO3D. 
The participants were asked to choose which of the two simulations (shown side-by-side) was more realistic. 
Here, LBS was preferred over CLO3D in \textbf{16.5\%} of the cases, HOOD in \textbf{16.9\%} of the cases, and \moniker{} in \textbf{35.8\%} of the cases.

These results demonstrate that, although \moniker{} cannot achieve the level of realism of commercial software such as CLO3D, its results are not much inferior in terms of perceived realism. 
For more details on the study please refer to the supplemental material.

\subsection{Automatic outfit resizing}
When designing 3D outfits, it is highly desirable to automatically adjust the size of the garments to the body shape.
However, in the case of multi-layer outfits, automatically resizing them may introduce unwanted cloth intersections.
To simulate such outfits, these penetrations have to be manually resolved by an artist because the simulation software may struggle to recover from them.

Since our method is trained to resolve existing intersections, it can handle penetrations caused by automatic resizing. 
We demonstrate this in Fig. \ref{fig:resize} and in the supplemental video.
Here, we use a single original outfit geometry that fits the canonical SMPL-X \cite{SMPL-X:2019} template body and then resize it to a new body shape by randomly sampling body shape parameters from a normal distribution with $\sigma=3$.
To resize the outfits, for each garment node, we sample shape blend shapes from the SMPL-X vertex closest to it.
We then modify the nodal positions using the same shape vectors as for the body.
Our method robustly recovers from the intersections that arise during this process, allowing it to realistically model the resized outfits.

%% file: include/figure_resize.tex
\begin{figure*}[ht]
  \centerline{  \includegraphics[width=\linewidth]{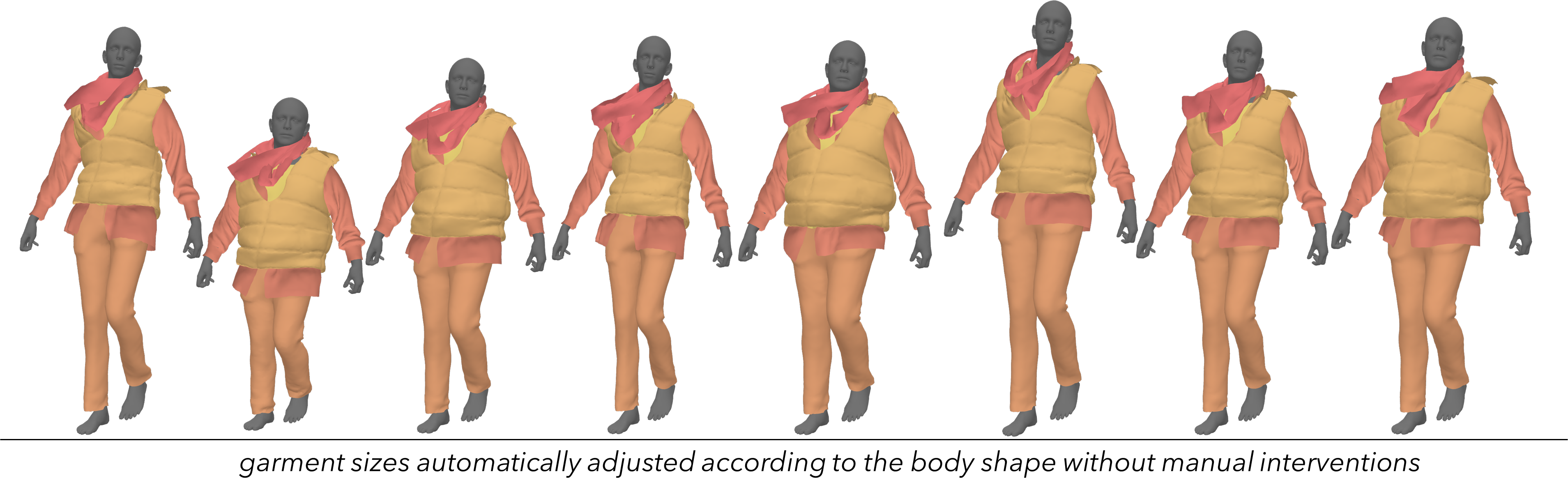}}
  \caption{
  Since our method can resolve intersections present in the initial geometry, we can model automatically resized outfits without manual resolution of the intersections arising during this process.
  }
  \label{fig:resize}
\end{figure*}

%% file: include/figure_penat50.tex
\begin{figure}[ht]
  \centerline{  \includegraphics[width=.8\linewidth]{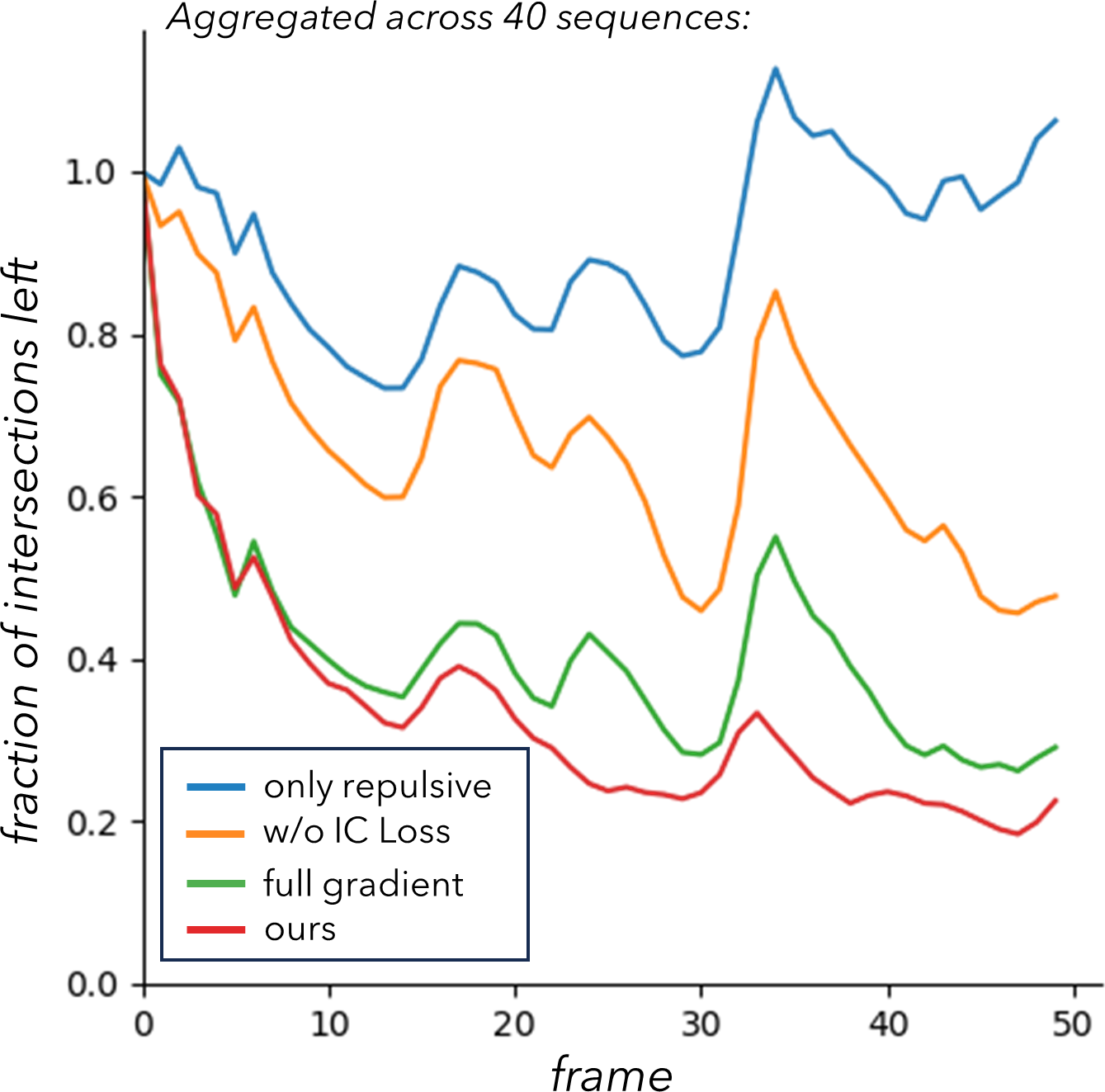}}
  \caption{The plot shows the fraction of the triangle-triangle intersections left after each frame (up to 50) relative to the initial geometry. The values are aggregated across the whole validation set (40 sequences). Note that during dynamic movements new intersections may appear, hence the plot is not monotonic.
    \vspace{-20pt}
  }
  \label{fig:penat50}
\end{figure}

%% file: include/table_statistics.tex
\begin{table}[]
\caption{We compare our method to three ablated versions in terms of an average number of penetrating triangle pairs. Each additional term provides  substantial improvements to collision handling.
\vspace{-10pt}
}
\resizebox{\columnwidth}{!}{%
\begin{tabular}{|l|c|c|c|c|}
\hline
                                     & 2 garments                                                     & 3 garments                                                     & 5 garments                                             & all outfits                                                    \\
                                     & \begin{tabular}[c]{@{}c@{}}avg. \\ 13016 \\ nodes\end{tabular} & \begin{tabular}[c]{@{}c@{}}avg. \\ 18981 \\ nodes\end{tabular} & \begin{tabular}[c]{@{}c@{}}36515 \\ nodes\end{tabular} & \begin{tabular}[c]{@{}c@{}}avg. \\ 20102 \\ nodes\end{tabular} \\ \hline
HOOD                                 & 1347                                                           & 5673                                                           & 38480                                                  & 8796                                                           \\ \hline
\multicolumn{1}{|c|}{only repulsive} & 373.2                                                          & 1069                                                           & 25220                                                  & 4424                                                           \\ \hline
w/o IC loss                          & 122.2                                                          & 268.4                                                          & 1491                                                   & 391                                                            \\ \hline
full gradient                        & 74.9                                                           & 163.4                                                          & 1315                                                   & 299                                                            \\ \hline
ours                                 & \textbf{55.9}                                                  & \textbf{126.6}                                                 & \textbf{481.2}                                         & \textbf{150}                                                   \\ \hline
\end{tabular}

}
\label{tab:ablation}
\vspace{-10pt}

\end{table}

%% file: 6_Conclusion.tex
\section{Conclusion}
ContourCraft is a novel method for modeling complex multi-layer outfits in motion.
Its core is a new Intersection Contour Loss term that allows the GNN-based model to resolve cloth intersections that are present in the initial geometry or occur during simulation.
At the same time, our model can prevent most of the penetrations from happening in the first place, ensuring realistic simulation.
In addition to the detailed quantitative analysis of our contributions, we show how to easily resize complex outfits to new body sizes while resolving interpenetrations. 
With the field of learned physical simulation of garments still in its infancy, we show that the flexibility of learned models holds promise for problems that are traditionally difficult for conventional physics-based simulation, including the resolution of existing self-intersections.

%% file: 7_Limitations.tex
\section{Limitations and Future Work}

Despite the overall efficiency of our  
learned method for handling cloth intersections, there are cases that are difficult to resolve due to the nature of the Intersection Contour loss.
For example, in dynamic sequences, small non-manifold pieces of cloth (e.g.~pockets) may pop outside the garment in a single frame.
In those cases, the model will assume that this is the correct configuration since trying to pull the pocket back inside would increase the length of the contour.
Similarly, if initialized with geometry in which two garments are entangled in a way where it is impossible to infer their correct order, our method will also fail.

While we base our method on a learned physical simulator ~\cite{hood}, one direction for future work is applying the formulation of Intersection Contour Loss to pose-driven garment deformation models such as PBNS~\cite{bertiche2020pbns} and SNUG~\cite{santesteban2022snug} that offer faster inference compared to HOOD.

\begin{acks}
AG was supported in part by the Max Planck ETH Center for Learning Systems. We thank Joachim Tesch, Christoph Gebhardt, Xu Chen, and Peter Kulits for their feedback and help during the project.
\end{acks}

%% file: include/figure_oneseqplot.tex
\begin{figure*}[ht]
  \centerline{  \includegraphics[width=\linewidth]{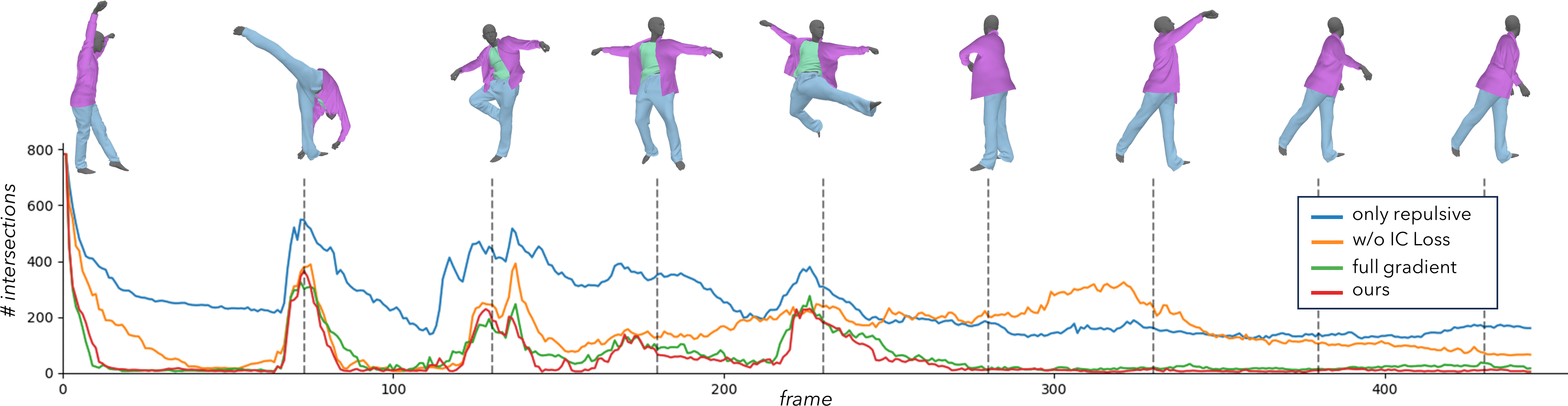}}
  \caption{
  For one of the validation sequences, we plot the number of intersecting triangle pairs for all compared ablations.
  Starting from an intersecting geometry, our method quickly resolves most penetrations. 
  During dynamic and complex motion sequences (for instance, those with body self-intersections), it may miss new penetrations but then is able to recover from them as well. 
  }
  \label{fig:oneseqplot}
\end{figure*}

%% file: 91_SM_Derivation.tex
\section{Intersection contour gradient derivation}

\definecolor{colord}{HTML}{007700}
\definecolor{colort}{HTML}{0000AA}
\newcommand{\gradd}[1]{\textcolor{colord}{#1}}
\newcommand{\gradt}[1]{\textcolor{colort}{#1}}

As shown in Figure 4 of the paper, we decompose the gradient of the Intersection Contour Loss into two components.
One component distorts the intersecting faces (green arrows), while another applies quasi-rigid translation. 

Here we show the derivation for both of these components in a simple case where a triangle $\Delta ABC$ penetrates an orthogonal plane with normal vector $\vec{n}$, as in Figure 4.

We derive the gradient of the loss function $\mathcal{L}_{IC}$ with respect to the position of the node $A$.
Thus we will only consider the penetration between the edge $AB$ and the plane. 
The derivation for edge $BC$ and other nodes follows the same steps.

Both edges of the triangle $\Delta ABC$ penetrate the plane. 
$AB$ penetrates it in the point $p_{I^0_i}$, while $BC$ in the point $p_{I^1_i}$.
We can compute the relative positions of the intersecting points $s_0$ and $s_1$ on each of these edges.

For $s_0$, we have
\begin{gather}
    s_0 = \frac{(x_{\Delta}\cdot \vec{n}) - (A\cdot\vec{n}) }{(B-A)\cdot\vec{n}}\ ,
    \label{eq:s0}
\end{gather}
where $x_\Delta$ is a point on the intersecting plane and $\vec{n}$ is the plane's normal.
We compute the position of the intersection point and the loss value as 
\begin{gather}
    p_{I^0} = A + s_0 (B-A) \\
    \mathcal{L}_{IC} = \| p_{I_i^0} - p_{I_i^1}\|^2 \label{eq:loss}
\end{gather}

From Eq.~\ref{eq:loss} we can derive the gradient of $\mathcal{L}_{IC}$ with respect to $p_{I_i^0}$ as
\begin{gather}
    \dfrac{\partial{\mathcal{L}_{IC}}}{\partial p_{I_i^0}} = 2(p_{I_i^0} - p_{I_i^1})\ .
\end{gather}
The gradient of $p_{I_i^0}$ w.r.t. $A$ comprises a \gradd{\textit{distortional}} and a \gradt{\textit{translational}} component:
\begin{gather}
    \dfrac{d p_{I_i^0}}{d A} = \gradd{\dfrac{\partial p_{I_i^0}}{\partial A}} + \gradt{\dfrac{\partial p_{I_i^0}}{\partial s_0} \dfrac{\partial s_0}{\partial A}}
\end{gather}
For the \gradd{\textit{distortional}} component, we have 
\begin{gather}
    \gradd{\dfrac{\partial p_{I_i^0}}{\partial A}} = (1 - s_0)I \ .
\end{gather}

The \gradt{\textit{translational}} component follows as
\begin{gather}
   \dfrac{\partial p_{I_i^0}}{\partial s_0} = B-A \\
   \dfrac{\partial s_0}{\partial A} = \frac{\vec{n}\cdot(x_\Delta-A)}{\vec{n}\cdot(B-A)}\vec{n} = k\vec{n}\label{eq:dsda},
\end{gather}

This is the derivative of $s_0$ over $A$ (see Eq.~\ref{eq:s0}).
For brevity, we substitute the scalar term in front of $\vec{n}$ with $k$.

We can now write the full \gradt{\textit{translational}} component as
\begin{gather}
    \gradt{\dfrac{\partial p_{I_i^0}}{\partial s_0} \dfrac{\partial s_0}{\partial A}} = k(B-A)\vec{n}^T \ .
\end{gather}

Finally, writing the full gradient of $\mathcal{L}_{IC}$ with respect to $A$, we again identify two components,
\begin{align}
    \dfrac{d{\mathcal{L}_{IC}}}{dA} & =
    \frac{d\mathcal{L}_{IC}}{dp_{I_i^0}} \frac{dp_{I_i^0}}{dA} \\
    & = \gradd{\frac{d\mathcal{L}_{IC}}{dp_{I_i^0}}\dfrac{\partial p_{I_i^0}}{\partial A}} + 
    \gradt{\frac{d\mathcal{L}_{IC}}{dp_{I_i^0}}\dfrac{\partial p_{I_i^0}}{\partial s_0} \dfrac{\partial s_0}{\partial A}} \\
    & = \gradd{2(1-s_0)(p_{I_i^0} - p_{I_i^1})} + \gradt{2k((p_{I_i^0} - p_{I_i^1})^T(B-A))\vec{n}}\ .
\end{align}
It is evident that the direction of the \gradd{\textit{distortional}} component is $p_{I_i^0} - p_{I_i^1}$, which is \gradd{\textit{parallel}} to the intersecting plane.
The direction of the \gradt{\textit{translational}} component is $\vec{n}$, which is \gradt{\textit{orthogonal}} to the intersecting plane.
To supervise our model we only use the \gradt{\textit{translational}} component.

%% file: 92_outfits.tex
\section{Experiment details}

\input{include/table_speed}

\subsection{Validation outfits}
For our experiments, we use 5 multi-garment outfits from BEDLAM~\cite{bedlam} dataset.
We downsample some of them to have a smaller number of nodes and fit on a single GPU during inference.

Table~\ref{tab:speed} shows the information on each outfit including the number of garments, total number of nodes, and average inference speed on validation sequences. 

\subsection{Perceptual study}
Here we provide a detailed description of the perceptual study presented in Section 5.3 of the main paper. 
We compare four methods: \moniker{}, HOOD~\cite{hood}, linear blend skinning (LBS), and the commercial software CLO3D~\shortcite{clo3d}.
HOOD simulates garments without considering interactions between pieces of cloth, which results in many overlaps and intersections within and between garments. 
LBS rigidly attaches clothing to the nearest part of the body, which doesn’t capture the natural movement of loose clothing very well.

For the study, we simulated eight different pose sequences and body shapes, consistent across all methods. 
Each method produced eight videos of the simulations, all with the same camera settings.

In the first part of the study, the participants were shown individual simulation videos and asked to rate them from 1 to 5.
The specific question asked was: \textit{"Mark how much do you agree with the following statement: \guillemotleft The motions of the clothing in the video are realistic and closely resemble how similar clothing would move in real life\guillemotright"}. 
The participants had to choose out of 5 options: \textit{"1 (completely disagree)"}, \textit{"2 (disagree)"}, \textit{"3 (neither agree nor disagree)"}, \textit{"4 (agree)"} or \textit{"5 (completely agree)"}.

The order in which the videos were presented was randomized. 
Each video received ratings from 22 different participants. 
Given that there were 8 sequences for each method, each method was evaluated 176 times in total.

In the second part of our study, we performed a direct side-by-side comparison between three pairs of methods, where each pair included CLO3D and one of the other simulation methods. 
Participants were presented with two videos at a time---one from CLO3D and one from another method---and were asked to select which video demonstrates a more realistic simulation.
The exact formulation was: \textit{In this task, you will see two animations of a character in which the motion of the clothing is digitally simulated.
Your task is to choose which clothing simulation is more realistic.
Please pay attention to the behavior of the clothing and not its colors. Ignore the way the body moves.}
The sequence of the video pairs and the order of the videos within each pair were randomized. Each pair of videos was evaluated by 37 different participants, resulting in each method pair being compared 296 times in total

%% file: include/table_speed.tex
\begin{table}[]
\begin{tabular}{|l|l|l|l|}
\hline
outfit name          & garments & nodes & avg. fps \\ \hline
\textit{cindy\_020}  & 2        & 12030 & 10.49    \\ \hline
\textit{caren\_008}  & 2        & 14003 & 10.18    \\ \hline
\textit{aaron\_022}  & 3        & 17745 & 9.33     \\ \hline
\textit{celina\_002} & 3        & 20218 & 6.2      \\ \hline
\textit{ben\_004}    & 5        & 36515 & 3.66     \\ \hline
\end{tabular}
\caption{ Number of garments and garment nodes in each of the five validation outfits from BEDLAM~\cite{bedlam}. We also list the average inference speed for each outfit. We run simulations with a time step of 1/30s. The timings were obtained using a single NVIDIA GeForce RTX 4090 GPU.}
\label{tab:speed}
\end{table}

%% file: 93_implementation.tex
\section{Implementation Details}

\subsection{Building input graph}
The model's input graph consists of the nodes of the outfit mesh $V^G$, nodes of the body mesh $V^B$, and several sets of edges.

We start by initializing the graph with nodes $V^G$ and edges $E^G$ of the garment. Then, we expand this by adding two levels of coarse edges $E^{C_i}$ (with $i$ indicating the level index). In the paper, these edges are included in $E^G$ for simplicity. For a detailed explanation of how these edges are constructed, please see the Supplementary Material of HOOD~\shortcite{hood}.

Next, we identify the nearest body node for each garment node. 
If they are closer than a threshold distance of $\epsilon = 3cm$, we create a body edge between them.
Finally, we follow Algorithm 2 of the main paper to add two sets of 'world edges' between the garment nodes that are nearby in world coordinates.
These edges are categorized as 'repulsive' $E^W_R$ and 'non-repulsive' $E^W_{NR}$, based on their positions relative to intersection contours.

In the end, the input graph can be represented as:

$$
G = \{V^G, E^G, V^B, E^B, E^{C_i}, E^W_R, E^W_{NR}\}
$$
Each node and edge in this graph is assigned a feature vector that includes physical data and additional parameters. We normalize these feature vectors to ensure that the distribution of each parameter approximates a normal distribution $\mathcal{N}(0, I)$.
This helps the model to converge. 
For more information on this process, refer to MeshGraphNets~\shortcite{pfaff2020learning}.



\subsection{Network architecture}
Our model follows the architecture used in HOOD~\shortcite{hood}, which includes an encoder, 15 message-passing steps, and a decoder. Each component consists of multiple multi-layer perceptrons (MLPs) with 2 hidden layers each and a latent dimension of 128. We use ReLU activation functions and layer normalization for these hidden layers, but no activation function in the output layer.

The \textbf{encoder} processes the input feature vectors of all nodes and edges to map them into a latent space. Each type of edge and the nodes have their dedicated MLP. These encoder MLPs output vectors of dimension 128, with the input dimension corresponding to the number of parameters each node or edge has.

Each \textbf{message-passing} step also contains MLPs for each edge set and an MLP for nodes. 
It first updates the latent vectors of each edge. This is done by concatenating the latent vector of each edge with the latent vectors of the nodes it connects to, and then feeding this combined vector into the corresponding MLP. The input dimension for these edge-specific MLPs is $128 \times 3$, and they output vectors of dimension 128.
Next, the step updates nodal latent vectors. 
This involves first aggregating the latent vectors from all the edges connected to a node, done separately for each set of edges. These aggregated vectors, along with the latent vector of the node itself, are then concatenated and processed through another MLP to generate the updated nodal vector. 
The input dimension for these nodal MLPs is $128 \times 6$, where $6$ is the number of edge sets and an output dimension of 128. 

For more details on the message-passing process please refer to MeshGraphNets~\shortcite{pfaff2020learning} and the Supp. Mat. of HOOD~\shortcite{hood}.

Finally, the \textbf{decoder} uses its MLP to transform the latent vectors of the garment nodes, each of size 128, into nodal acceleration vectors of size 3. 
These acceleration vectors are subsequently de-normalized using pseudo-statistics derived from sequences generated by linear blend skinning. For more detailed information on this process, please refer to MeshGraphNets~\shortcite{pfaff2020learning} and the Supp. Mat. of HOOD~\shortcite{hood}.

\subsection{Autoregressive training}
Following HOOD, we gradually increase the number of autoregressive steps in each training sample from 1 to 5 every 5000 iterations. 
In the second and third stages, we also include full-length pose sequences of up to 150 frames each.
In these stages, we also alternate between training samples with and without cloth--cloth interactions.
For that, in every second training iteration, we omit cloth--cloth correspondences and the respective objective terms, falling back to the process from the first stage.
This helps the model avoid diverging from generating realistic collision-agnostic behavior of the fabric.